\documentclass[aps,preprint,showpacs]{revtex4}

\usepackage{graphicx}
\usepackage{bm}
\begin{document}
\setcounter{page}{1}
\title{Coherent control of spontaneous emission of a three-level atom in a coherent photonic band gap reservoir}
\author{Szu-Cheng \surname{Cheng$^{1}$}}
\email {sccheng@faculty.pccu.edu.tw}
\thanks{FAX: +886-2-28610577}
\author{Jing-Nuo \surname{Wu$^{2}$}}
\email {jingnuowu@so-net.net.tw}
\thanks{FAX: +886-5-2717909}
\author{Tzong-Jer \surname{Yang$^{3}$}}
\author{Wen-Feng \surname{Hsieh$^{4}$}}
\affiliation{$^{1}$Department of Physics, Chinese Culture University, Taipei, Taiwan, R. O. C.}
\affiliation{$^{2}$Department of Applied Physics, National Chiayi University, Chiayi, Taiwan, R. O. C.}
\affiliation{$^{3}$Department of Electrical Engineering, Chung Hua University, Hsinchu, Taiwan, R. O. C.}
\affiliation{$^{4}$Department of Photonics, National Chiao Tung University, Hsinchu, Taiwan, R. O. C.}
\date[]{Received \today }
\begin{abstract}
By studying the fluorescence and optical properties of a three-level system, we propose a new point of view on the coherent control of these spectra. With the definite phase difference between the fields of the air band and dielectric band in photonic band gap (PBG) reservoirs, the spectra of spontaneous emission, absorption, and dispersion exhibit the coherent property and quantum interference effect. This coherent interference depending on the position of the embedded atom and the width of band gap causes the coupling of the free-space light and the PBG light to result in blue shift of spectra and the appearance of dark lines and kinks. By coherently controlling the position-dependent dispersion, we can tune the frequency of slow light.   
\end{abstract}

\pacs{42.50.Gy, 42.70.Qs}

\maketitle



It is well-known that spontaneous emission (SE) rate and optical properties can be modified effectively by placing the atoms in photonic band gap (PBG) materials \citep{RFNabiev93,SJohn94, AGKofman94, TQuang97, SYZhu97, YYang99, SBay96, SBay97a, NVats98}, where the density of modes of the reservoirs has significant deviation from that of free space vacuum. This modification changes the atomic coherence and quantum interference effects and provides potential application to quantum optical communication. The related atomic coherence effects include the electromagnetically-induced transparency (EIT) \citep{SEHarris97}, lasing without inversion \citep{SEHarris99, SEHarris89}, slow propagation of light \citep{CLiu01}, and nonlinear effects at low light level \citep{MOScully89}.

Here we propose a new point of view on controlling the SE and optical resonant spectra through changing the relative position of embedded atom in photonic crystals (PCs). It originates from the fact that the strength of photon-atom interaction depends on the atomic position \citep{NVats02} and two band reservoirs become coherent and can interfere with each other, which results from the definite phase difference between the air-band and dielectric-band fields of PBG \citep{Joannopoulos08}. One band approximation is valid when the PBG is relatively wide and the transition frequencies of the atoms is near the air or dielectric band edge. For the photonic crystals with narrow band gap, the photon-atom coupling of the embedded atom system has to consider the contribution of both electric fields from the air-band and dielectric-band reservoirs. Existence of both reservoirs leads to a stronger photon-atom coupling \citep{NVats02}, and the time evolution decay of the excited population of a two-level atom has been shown faster than that in the one band approximation. Although atomic coherence in a $\Lambda$-type three-level atom embedded in a PBG structure has been studied using two band model \citep{EPaspalakis99a, EPaspalakis99b, EPaspalakis01}, it was assumed that the atom inside PC interacted with two incoherent PBG reservoirs. The results of Ref.\onlinecite {EPaspalakis99a, EPaspalakis99b, EPaspalakis01} revealed that the photon-atom coupling strengths of both reservoirs were the same and independent of the position of embedded atom. In this letter, a $\Lambda$-type three-level atom embedded in a PBG structure is studied (Fig.1). One transition level of the atom ($\left\vert2\right\rangle\leftrightarrow\left\langle1\right\vert$) lies near the PBG edge; the other transition ($\left\vert2\right\rangle \leftrightarrow\left\langle0\right\vert $) is far from the PBG edge and couples with the Markovian reservoir. The most interesting contains the (long time) spontaneous emission spectrum in the Markovian reservoir and strength of coherent coupling with interference effect, which depends on the embedded position of the atom and the width of band gap. The variation of the atomic position and band-gap width would lead to the change of the SE rate and the spectra of absorption and dispersion. These results provide a new degree of freedom to control the SE and optical properties in PCs, especially for nonlinear optical phenomena. 
 
Because the vector potential in the interaction Hamiltonian of the system is a function of atom position in a Wigner-Seitz cell $\vec{r}_{0}$ \citep{NVats02}, under the rotating-wave and electronic dipole approximations, we modified the interaction Hamiltonian of Ref. \onlinecite{EPaspalakis99a} to be $H=\hbar\sum_{\lambda}g_{\lambda}e^{-i(\omega_{\lambda}-\omega_{20})t}\left\vert 2\right\rangle \left\langle 0\right\vert
\hat{a_{\lambda}}+\hbar\sum_{a}g_{a}(\vec{r}_{0})e^{-i(\omega_{a}-\omega_{21})t}\left\vert 2\right\rangle \left\langle 1\right\vert \hat{a_{a}}+\hbar\sum_{d}g_{d}(\vec{r}_{0})e^{-i(\omega_{d}-\omega_{21})t}\left\vert 2\right\rangle \left\langle 1\right\vert \hat{a_{d}}+H.c.$ Here $g_{a(d)}(\vec{r}_{0})=\frac{\omega_{21}}{\hbar}d_{21}\left( \frac{\hbar}{2\epsilon_{0}\omega_{a(d)}V}\right)^{1/2}\widehat{d}_{21}\cdot\overrightarrow{E}_{a(d)}^{\ast}(\vec{r}_{0})$ characterizes the coupling strength of the atom with air(dielectric)-band-reservoir modes $a$($d$) and is dependent on the atomic position $\vec{r}_{0}$. The physical origin of the PBG \citep {Joannopoulos08} arises from where the field energy is located -- the air band being concentrated in the low dielectric layers and the dielectric band in the high dielectric layers. The distributions of the fields $\overrightarrow{E}_{a}^{\ast}(\vec{r}_{0})$ and $\overrightarrow{E}_{d}^{\ast}(\vec{r}_{0})$ from two-band reservoirs change coherently with $\pi$/2 phase difference and can be expressed as $\overrightarrow{E}_{a,\vec{k}}^{\ast}=E_{\vec{k}}\cos\theta(\vec{r}_{0})\hat{\epsilon}$ and $\overrightarrow{E}_{d,\vec{k}}^{\ast}=E_{\vec{k}}\sin\theta(\vec{r}_{0})\hat{\epsilon}$ ($\hat{\epsilon}$ is a unit vector of the electric field). Here these two coherent eigenmode fields contain only a single amplitude $E_{\vec{k}}$. And the angle parameter $\theta(\vec{r}_{0})$ is a function of atomic position $\vec{r}_{0}$. Thus, the atomic position-dependent coupling strength is $g_{a}(\vec{r}_{0})=g_{\vec{k}}\cos\theta(\vec{r}_{0})$ ($g_{d}(\vec{r}_{0})=g_{\vec{k}}\sin\theta(\vec{r}_{0})$) with the real constant $g_{\vec{k}}=\frac{\omega_{21}}{\hbar}\left(\frac{\hbar}{2\epsilon_{0}\omega_{\vec{k}}V}\right)^{1/2} E_{\vec{k}}(\overrightarrow d_{21}\cdot\hat{\epsilon})$. We denoted the atomic transition frequency as $\omega_{ij}$, photonic eigen-mode frequency as $\omega_{\vec{k}}$, and fixed polarization orientation of atomic dipole moment as $\overrightarrow{d_{21}}$ \citep{YSZhou06}.

As the SE rate is studied, we assume the atom is initially excited in state $\left\vert 2\right\rangle$. By substituting the interaction Hamiltonian and single-photon wave function into the time-dependent Schr$\ddot{o}$dinger equation, we have the time evolution of the probability amplitudes    
\begin{eqnarray}
i\frac{d}{dt}\alpha_{2}(t) &  =\sum_{\lambda}
g_{\lambda}\alpha_{0\lambda}(t)e^{-i(\omega_{\lambda}-\omega_{20}
)t}+\sum_{a}g_{a}(\vec{r}_{0})\alpha_{1a}(\vec{r}_{0},t)e^{-i(\omega_{a
}-\omega_{21})t}\nonumber\\
&  +\sum_{d}g_{d}(\vec{r}_{0})\alpha_{1d}(\vec{r}_{0},t)e^{-i(\omega_{d
}-\omega_{21})t},
\end{eqnarray}

\begin{equation}
i\frac{d}{dt}\alpha_{0\lambda}(t) =g_{\lambda}\alpha_{2}(t)e^{i(\omega_{\lambda}-\omega_{20})t},
\end{equation}
\begin{equation}
i\frac{d}{dt}\alpha_{1a}(\vec{r}_{0},t) =g_{a}(\vec{r}_{0})\alpha_{2}(t)e^{i(\omega_{a}-\omega_{21})t},
\end{equation}

\begin{equation}
i\frac{d}{dt}\alpha_{1d}(\vec{r}_{0},t) =g_{d}(\vec{r}_{0})\alpha_{2}(t)e^{i(\omega_{d}-\omega_{21})t}.
\end{equation}
Here $\alpha_{2}(t)$ and $\alpha_{0\lambda}(t)$ are the probability amplitudes of the system in the states $\left\vert2,\{0_{\lambda},0_{a},0_{d}\}\right\rangle$ and $\left\vert 0,\{1_{\lambda},0_{a},0_{d}\}\right\rangle$, which describe the atom in the state $\left\vert 2\right\rangle$ with no photon and in the state $\left\vert 0\right\rangle$ with a single photon in the mode $\lambda$, respectively. 

The probability amplitudes of the system in the states $\left\vert 1,\{0_{\lambda},1_{a},0_{d}\}\right\rangle$ and $\left\vert 1,\{0_{\lambda},0_{a},1_{d}\}\right\rangle$ (the atom in the state $\left\vert 1\right\rangle$ and a photon in either the air or dielectric band) are considered to be functions of atomic position  and assumed to be $\alpha_{1a}(\vec{r}_{0},t)=\alpha_{1k}(t)\cos\theta(\vec{r}_{0})$ and $\alpha_{1d}(\vec{r}_{0},t)=\alpha_{1k}(t)\sin\theta(\vec{r}_{0})$, respectively. 

Our aim is to derive the long time spontaneous emission spectrum of the transition $\left\vert2\right\rangle\leftrightarrow\left\langle0\right\vert$ within the Markovian reservoir, namely, ${S(\delta_{\lambda})}\propto{\left\vert{\alpha_{0\lambda}(t\rightarrow\infty)}\right\vert}^{2}$ with the detuning frequency $\delta_{\lambda}=\omega_{\lambda}-\omega_{20}$. Through performing Laplace transform and the final-value theorem, we have $S(\delta_{\lambda})\propto\gamma\left\vert\widetilde{\alpha}_{2}(s=-i\delta _{\lambda})\right\vert ^{2}$ with $\widetilde{\alpha}_{2}(s)$ being the Laplace transform of the probability amplitude $\alpha_{2}(t)$, which is given by 

\begin{eqnarray}
\widetilde{\alpha}_{2}(s)=\left\{1-\sin^{2}\theta(\vec{r}_{0})\cos^{2}\theta(\vec{r}_{0})\left[
\widetilde{\alpha}_{2}(s+iD_{\omega})\widetilde{K}_{1}(s+iD_{\omega})
+\widetilde{\alpha}_{2}(s-iD_{\omega})\widetilde{K}_{2}(s-iD_{\omega})\right] \right\}\nonumber\\  
\linebreak\ast \left[  s+\gamma/2+\cos^{4}\theta(\vec{r}_{0})\widetilde{K}_{2}(s)+\sin^{4}\theta(\vec{r}_{0})\widetilde{K}_{1}(s)\right]^{-1}.  
\end{eqnarray}
   
Here $D_{\omega}=\omega_{a}-\omega_{d}$ is the frequency difference of the air-band and dielectric-band modes at a fixed wave vector which could be approximated by the band gap $\Delta=\omega_{c}-\omega_{v}$ because most of the PBG density of states (DOS) is contributed from the states near band edges (see Fig.(c)). Besides, we have applied the Weisskopf-Wigner result \citep{SJohn95} to the free space modes $\lambda$ and the dispersion relation of a two-band isotropic effective mass model to non-Markovian memory kernels $K_{1}(t)=\beta^{3/2}e^{i\pi/4}e^{-i\delta_{v}t}/\sqrt{t}$ and $K_{2}(t)=\beta^{3/2}e^{-i\pi/4}e^{-i\delta_{c}t}/\sqrt{t}$. The Laplace transforms of these non-Markovian kernels are $\widetilde{K}_{1}(s)=\beta^{3/2}e^{i\pi/4}/(2\sqrt{s+i\delta_{v}})$ and $\widetilde{K}_{2}(s)=\beta^{3/2}e^{-i\pi/4}/(2\sqrt{s+i\delta_{c}})$, which have two singularities exhibiting two dark lines of the SE spectrum \citep{EPaspalakis99a, EPaspalakis99b, EPaspalakis01}. We used the iterative method once in this nonlocal difference equation $\widetilde{\alpha}_{2}(s)$ to get the SE spectrum. From the structure of $\widetilde{\alpha}_{2}(s)$, we observe that the shape of the SE spectrum is intimately controlled by the factors of atomic position and quantum interference effects. 

The SE spectra were plotted as a function of detuning frequency $\delta_{\lambda}=\omega-\omega_{20}$ for several gap  widths ($\Delta$) in Fig. 2. The symmetric case in which the atomic transition frequency $\omega_{21}$\ is chosen at the middle of the gaps is shown in Fig. 2(a). As expected for the large gape width $\Delta=6$, the excited atoms mainly couple to the free-space vacuum and thus the spectrum shows a single Lorentzian peak \citep{EPaspalakis99a, EPaspalakis99b, EPaspalakis01}, which is referred as the free space light. However, as decreasing the gap width, other than the main peak, we observed two symmetric side lobes separated by zeros, which result from strong coupling of the excited atoms through the $\left|2\right\rangle\rightarrow\left|1\right\rangle$ transition with the PBG reservoir at these two band edges, where have the largest DOS. The zeros are also termed the dark lines \citep{EPaspalakis99a, EPaspalakis99b, EPaspalakis01}. These two side lobes originate from the atomic free-space transition coupling to the PBG vacuum, which are referred as the PBG light. This coherent coupling effect grows stronger for the smaller gap widths. For the asymmetric case, in which the atomic transition frequency $\omega_{21}$ is located at the air-band edge, the spectrum of the larger PBG is close to that of the single-band case \citep{SJohn94}; whereas, as decreasing the gap width, the free-space light will be significantly quenched by the atom's emitting the PBG light or loss population through $\left|2\right\rangle\rightarrow\left|1\right\rangle$ transition. These results dramatically differ from those of the previous model \citep{EPaspalakis99a, EPaspalakis99b, EPaspalakis01}, which considered the non-Markovian reservoir as two incoherent PBG reservoirs and showed no intensity quench. Our model reveals, for the smaller gap width, apparent quantum interference effect (quench in the main peak and enhance dark lines at the band edges) as a result of considering the fields of the non-Markovian reservoirs as coherent mode fields. The interference of the free-space light with PBG light from air band is the same as that from dielectric band for the symmetric case ; whereas, it is enhanced in the air band for the asymmetric case ($\omega_{21}$ at the air-band edge). Besides, this quantum interference effect also happens when the field modes from the air band interfere strongly with the modes from the dielectric band via the atom. Its result is shown by two shallow kinks of two side lobes for symmetric case and by abrupt kink of the right-hand side lobe for the asymmetric case in small gap width $\Delta=1$. 

Next, we shall consider how the SE spectrum is affected by the relative position of the atom in a
Wigner-Seitz cell. Both the symmetric case and asymmetric case were considered for narrow gap ($\Delta=1$) and shown in Fig. 3(a) and Fig. 3(b). As increasing the $\theta(\vec{r}_{0})$ value, we observed that for both the symmetric and asymmetric cases, the free-space light makes a blue shift with increasing $\theta(\vec{r}_{0})$ but does not have much change in intensities. The increase of $\theta(\vec{r}_{0})$ causes the growth of the dielectric-band field strength ($\overrightarrow{E}_{d,\vec{k}}^{\ast}=E_{\vec{k}}\sin\theta(\vec{r}_{0})\hat{\epsilon}$) and thus the stronger coherent coupling with dielectric-band field to push the free-space light toward the air band. Therefore, the interference between the free-space light and the air-band PBG light is then enhanced. It results in the blue shift and increasing radiated power of the air-band PBG light. For the asymmetric case ($\omega_{21}$ at the air-band edge), the resonant effect enhances the more radiated power of the air-band PBG light such that the radiated power of the air-band PBG light eventually becomes stronger than that of the free-space light at $\theta(\vec{r}_{0})=\pi/3$. 

When the absorption and dispersion properties of the system are studied, we apply a weak probe laser \citep {CGDu02} with field angular frequency $\omega\cong\omega_{20}$ to the atomic initial ground state $\left\vert 0\right\rangle$. With the same consideration of the coherent reservoir of the system, we obtained the time evolution of the probability amplitudes 
\begin{equation}
i\frac{d}{dt}b_{0}(t)=\Omega b_{2}(t),
\end{equation}
\begin{eqnarray}
i\frac{d}{dt}b_{2}(t) &  =\Omega b_{0}(t)-(\delta+i\frac{\gamma}{2}%
)b_{2}(t)+e^{i\delta t}\sum_{k}g_{k}b_{1k}(t)e^{-i(\omega
_{a}-\omega_{21})t}\cos^{2}\theta(\vec{r}_{0})\nonumber\\
& +e^{i\delta t}\sum_{k}g_{k}b_{1k}(t)e^{-i(\omega
_{d}-\omega_{21})t}\sin^{2}\theta(\vec{r}_{0}),
\end{eqnarray}
\begin{equation}
i\frac{d}{dt}b_{1k}(t)   =e^{-i\delta t}g_{k}b_{2}(t)e^{i(\omega_{a
}-\omega_{21})t}\cos^{2}\theta(\vec{r}_{0}) +e^{-i\delta t}g_{k}b_{2}(t)e^{i(\omega_{d}-\omega_{21})t}\sin
^{2}\theta(\vec{r}_{0}).
\end{equation}

Here we have added a free-space decay rate $\gamma$ to the excited state $\left\vert 2\right\rangle $ and applied $g_{a}(\vec{r}_{0})=g_{\vec{k}}\cos\theta(\vec{r}_{0})$ ($g_{d}(\vec{r}_{0})=g_{\vec{k}}\sin\theta(\vec{r}_{0}$)) and $b_{1a}(\vec{r}_{0},t)=b_{1\vec{k}}(t)\cos\theta(\vec{r}_{0})$ ($b_{1d}(\vec{r}_{0},t)=b_{1\vec{k}}(t)\sin\theta(\vec{r}_{0})$) to the position-dependent coupling strength and probability amplitudes. The probability amplitudes $b_{2}(t)$, $b_{0}(t)$, and $b_{1a}(\vec{r}_{0},t)$ ($b_{1d}(\vec{r}_{0},t)$) express the system in 'bare' states $\left\vert 2,\{0_{a
},0_{d}\}\right\rangle $, $\left\vert 0,\{0_{a
},0_{d}\}\right\rangle $, and $\left\vert 1,\{1_{a
},0_{d}\}\right\rangle $ ($\left\vert 1,\{0_{a
},1_{d}\}\right\rangle $) with initial conditions $b_{0}(t=0)=1$ and $b_{2}(t=0)=b_{1a}(t=0)=b_{1d}(t=0)=0$. The laser detuning frequency is denoted by $\delta=\omega-\omega_{20}$ and Rabi frequency by $\Omega=-\overrightarrow{d}_{20}\cdot\overrightarrow{E}$ with $\overrightarrow{E}$ being the electric field of the probe laser field. The absorption and dispersion properties can be obtained from the steady-state linear susceptibility $\chi$ of the system.  Under the assumption
of a weak laser-atom interaction and $\left\vert 0\right\rangle
\leftrightarrow\left\vert 2\right\rangle $ transition occurring
in Markovian reservoir, this linear susceptibility
can be expressed as \citep{EPaspalakis99a, EPaspalakis99b, EPaspalakis01}

\begin{equation}
\chi(\delta)=-\frac{4\pi\aleph\left\vert \overrightarrow{d}_{20}\right\vert
^{2}}{\Omega}b_{0}(t\rightarrow\infty)b_{2}^{\ast}(t\rightarrow\infty).
\end{equation}
We assume the atomic density-related constant $4\pi\aleph\left\vert
\overrightarrow{d}_{20}\right\vert ^{2}\equiv1.$, the Rabi frequency $\Omega<<\gamma$ (decay rate), and $b_{0}(t)\approx1$ for all times. Here the long-time behavior of the probability
amplitude $b_{2}(t\rightarrow\infty)$ could be resolved from the final-value theorem and the Laplace transform, so the linear susceptibility is
\begin{equation}
\chi(\delta)=-\frac{1}{\delta-i\gamma/2-i\cos^{4}\theta(\vec{r}_{0})\widetilde{K}_{2}^{\ast
}(-i\delta)-i\sin^{4}\theta(\vec{r}_{0})\widetilde{K}_{1}^{\ast}(-i\delta)}
\end{equation}
with $\widetilde{K}_{1}^{\ast}$ and $\widetilde{K}_{2}^{\ast}$ being the complex conjugates of the Laplace transforms of the non-Markovian kernels, which are determined by the PBG model. The absorption $(-Im[\chi(\delta)])$ and dispersion $(Re[\chi(\delta)])$ spectra
of the probe field are thus affected by the DOS of the PBG
reservoirs. Besides, these spectra are also affected by the atomic position inside
the PC because of the position-dependent angle
$\theta(\vec{r}_{0})$.

The absorption and dispersion spectra for two-coherent-bands system were plotted based on Eq.(10) in Fig. 4. The absorption spectra exhibits similar behavior to the spontaneous-emission ones having a central peak with two side lobes and blue shift in the central peak at increasing $\theta(\vec{r}_{0})$. The two-band absorption profiles of the atomic-position dependence is shown in Fig. 4(a) for the atomic resonance $\omega_{21}$ at the middle of the gap (symmetric case). The transparent windows are independent of the relative position of the atom ($\theta(\vec{r}_{0})$) because they are determined by the DOS of the PBG reservoirs \citep{EPaspalakis99a, EPaspalakis99b, EPaspalakis01}. As the field strength from dielectric band of the PBG reservoirs grows stronger ($\theta(\vec{r}_{0})$ is increased), the central peak of absorption is lying close to the air-band edge (right hand side). And the absorption on the air-band side lobe is enhanced while the dielectric-band side lobe is suppressed due to coherence coupling and quantum interference effect. As $\theta(\vec{r}_{0})$ increases, stronger coupling with the dielectric-band field pushes the free-space absorption line (central peak)  toward the air-band edge. Stronger quantum interference between this central-peak light and the PBG light of the air band enhances the air-band side lobe intensity. This quantum interference effect is due to the coherence of the PBG reservoirs, which has more enhanced effect for the smaller PBG. The coherent property can also affect the transmission of the probe light shown in Fig. 4(b). The frequency of the slow velocity photons shifts toward blue side for increased $\theta(\vec{r}_{0})$. That is, the slow photons with higher frequency represent the stronger coupling with the dielectric-band field.

In conclusion, we have studied the spectra of spontaneous emission, absorption and dispersion of a
three-level atom with $\Lambda$-configuration in a PBG reservoir using two-band isotropic effective-mass model. With the consideration of the definite phase difference between the air-band and dielectric-band fields, we found that two band reservoirs are coherent and can interfere with each other. The coupling strength of the atom with this coherent reservoir depends on the embedded position of the atom. This is quite different from the previous two-band studies, which took the PBG reservoir as two independent and incoherent reservoirs and assumed equal strength of coupling between the atom and each individual reservoir of the PBG. Our SE spectra show the coherence phenomenon of the PBG reservoirs through the intensity changes of the free-space light and PBG light. Besides, for the small PBG, the spectra of the PBG light reveal quantum interference effect between the PBG reservoirs by means of kinks of side lobes. The position-dependent coupling strength is shown by the blue shift of the free-space light in SE spectra, which could be verified by experiment. The profiles of the absorption and dispersion show the coherent and quantum interference properties, too. With the variation of $\theta(\vec{r}_{0})$, we could observe the coherent property of the reservoir by the shift of the free-space absorption line. The intensity changes of the absorption side lobe illustrate the quantum interference effect of the system with the small PBG. The coherent property could also be observed in the dispersion spectra, whose frequencies of the slow photons vary with parameter $\theta(\vec{r}_{0})$. It results in the change of the refraction index with the atomic position. 

We greatly acknowledge partial support from the National
Science Council of Republic of China under Contract Nos. NSC
96-2914-I-009-017, NSC 95-2119-M-009-029, NSC
96-2112-M-034-002-MY3, and NSC 96-2628-M-009-001.

\bibliography{PRL0504}
\begin{figure}
\includegraphics[width=13.0 cm]{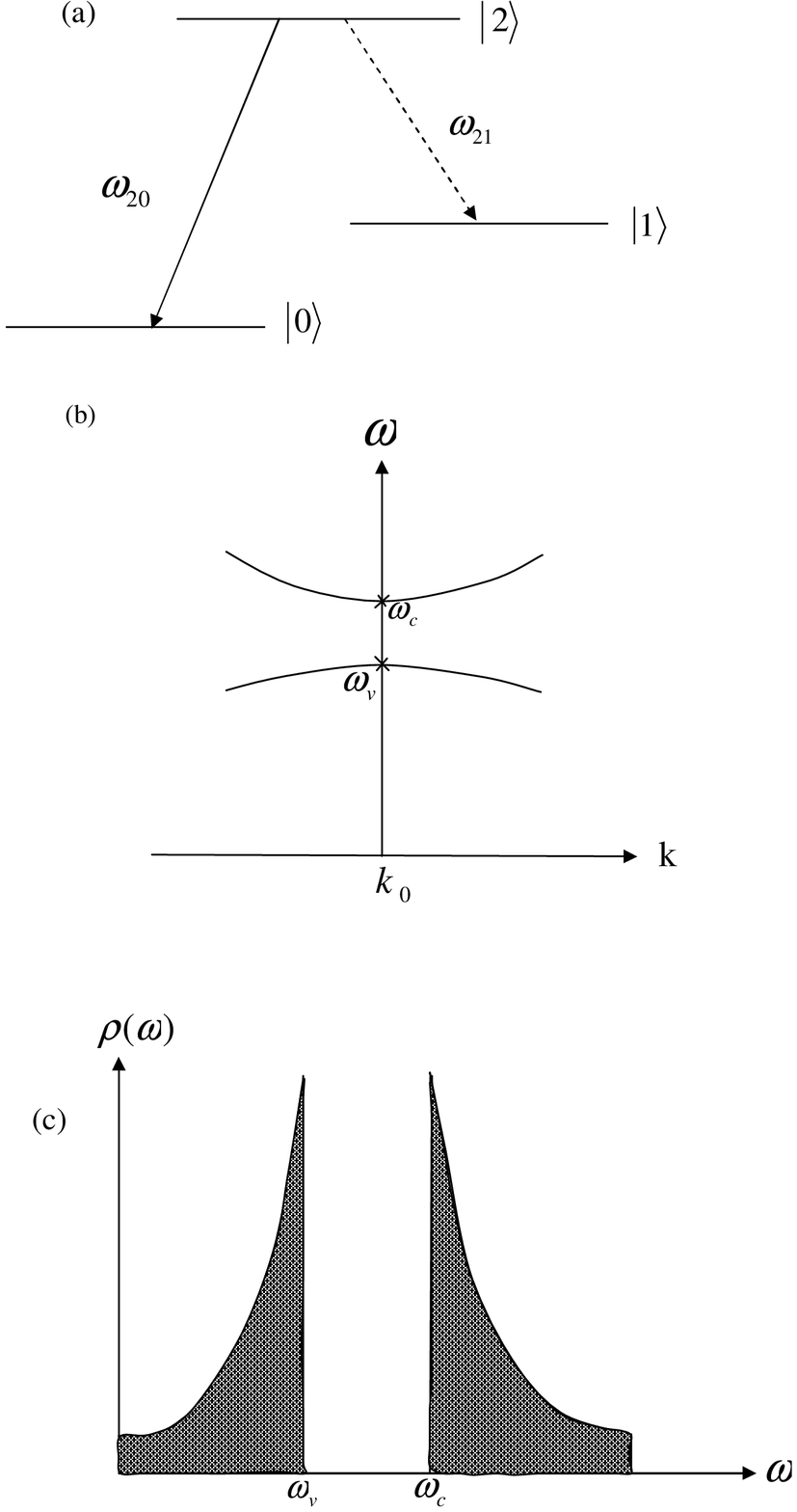} \label{Fig1}
\caption{(a) A $\Lambda$-type three-level atom with excited state $\left\vert2\right\rangle$, intermediate state $\left\vert1\right\rangle$, and ground state $\left\vert0\right\rangle$. The transition frequency $\omega_{21}$ lies near the PBG edges and $\omega_{20}$ far from the PBG edges. (b) Dispersion relation near the PBG edge with $\omega_{c}$ and $\omega_{v}$ being the air-band and dielectric-band edge frequencies. (c) Density of states (DOS) of the two-band isotropic effective mass model.}

\end{figure}

\begin{figure}
\includegraphics[width=12.0 cm]{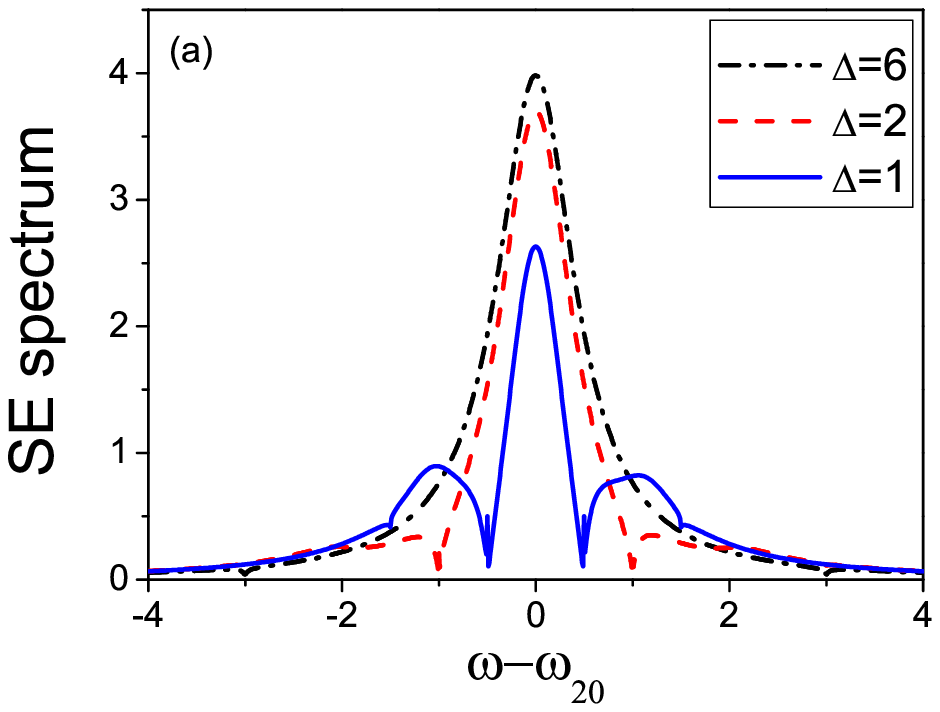}\label{Fig2a}
\includegraphics[width=12.0 cm]{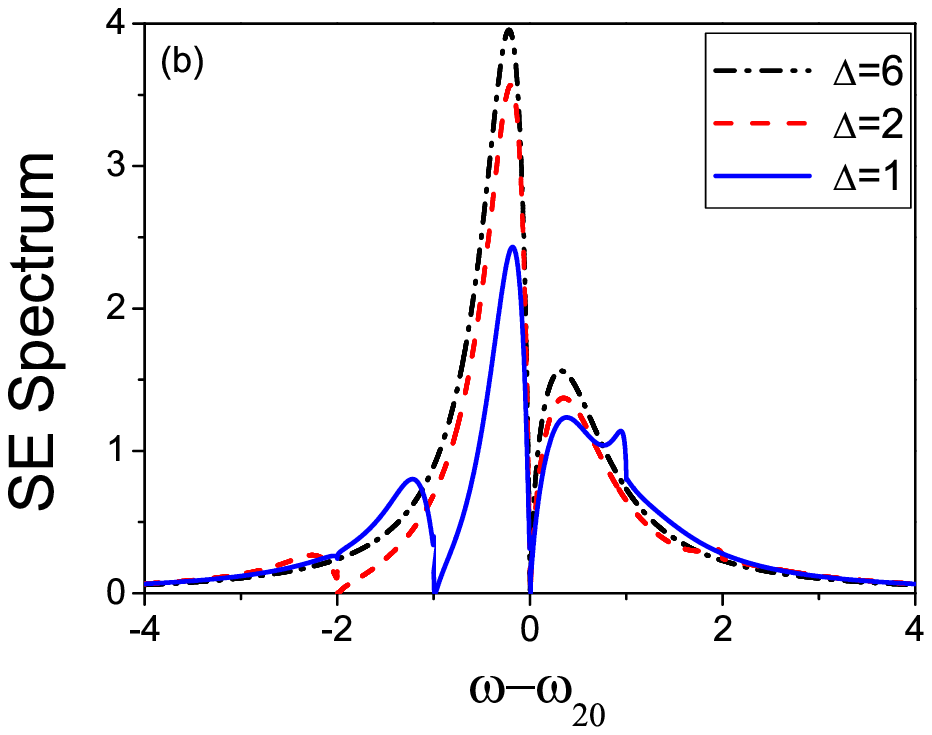}\label{Fig2b}
\caption{ The SE spectra for several gap widths ($\Delta$). The parameters (in units of coupling constant $\beta$) are decay rate $\gamma=1$, $\theta(\vec{r}_{0})$=$\pi$/4, PBG width $\Delta$=6 (dot-dashed line), $\Delta$=2 (dashed line), $\Delta$=1 (solid line), and (a) detuning frequency of band edge $\delta_{c}$=-$\delta_{v}$=$\Delta$/2 for the symmetric cases;
(b) $\delta_{c}$=0, $\delta_{v}$=-$\Delta$ for the asymmetric cases. ($\delta_{c}$=$\omega_{c}-\omega_{21}$, $\delta_{v}$=$\omega_{v}-\omega_{21}$)}
\end{figure}
\begin{figure}
\includegraphics[width=12.0 cm]{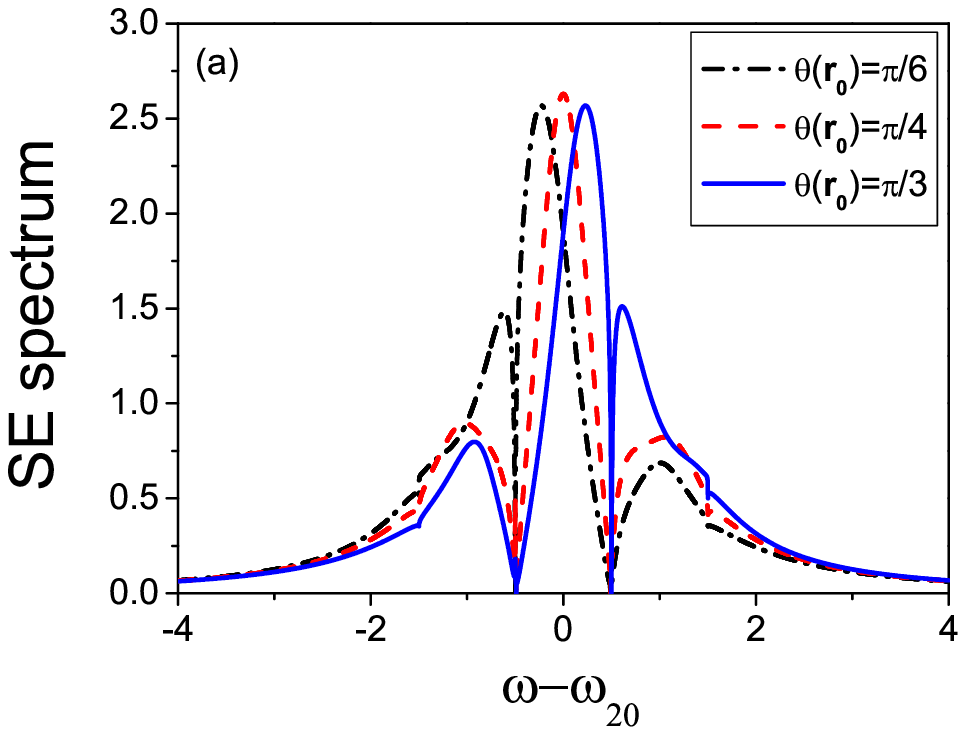} \label{Fig3a}
\includegraphics[width=12.0 cm]{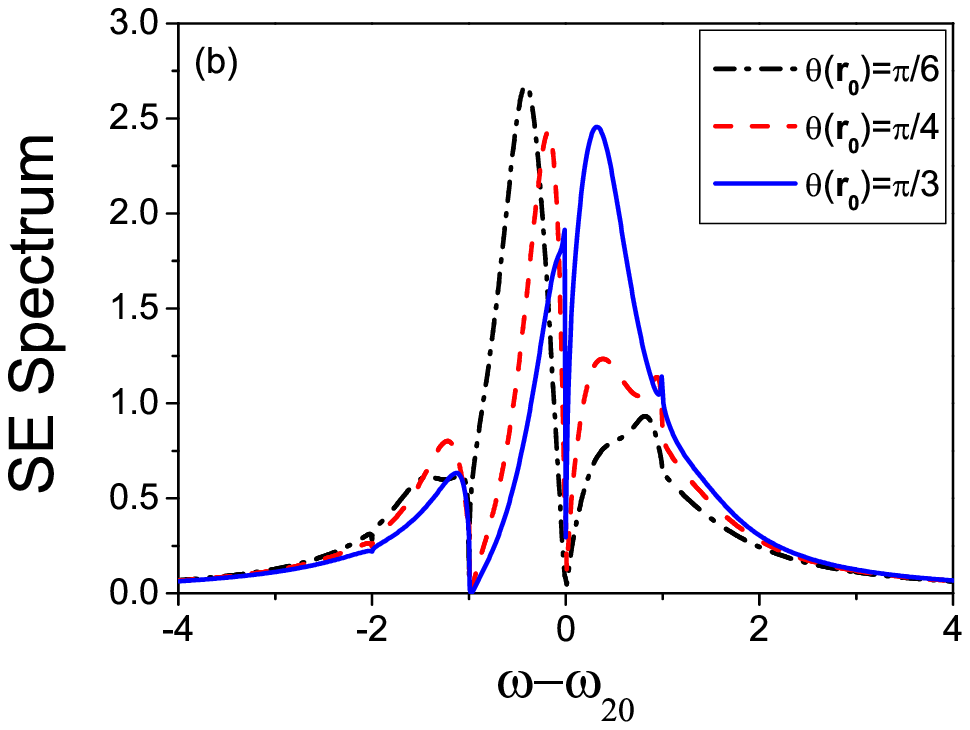} \label{Fig3b}
\caption{ The SE spectra for several position-dependent parameters $\theta(\vec{r}_{0})$. The parameters (in units of $\beta$) are $\gamma$=1,  $\Delta$=1, $\theta(\vec{r}_{0})$=$\pi$/6 (dot-dashed line), $\theta(\vec{r}_{0})$=$\pi$/4 (dashed line), $\theta(\vec{r}_{0})$=$\pi$/3 (solid line), and (a)$\delta_{c}$=-$\delta_{v}$=0.5 for the symmetric cases; (b)$\delta_{c}$=0, $\delta_{v}$=-1 for the asymmetric cases. ($\delta_{c}$=$\omega_{c}-\omega_{21}$, $\delta_{v}$=$\omega_{v}-\omega_{21}$)}

\end{figure}


\begin{figure}
\includegraphics[width=12.0 cm]{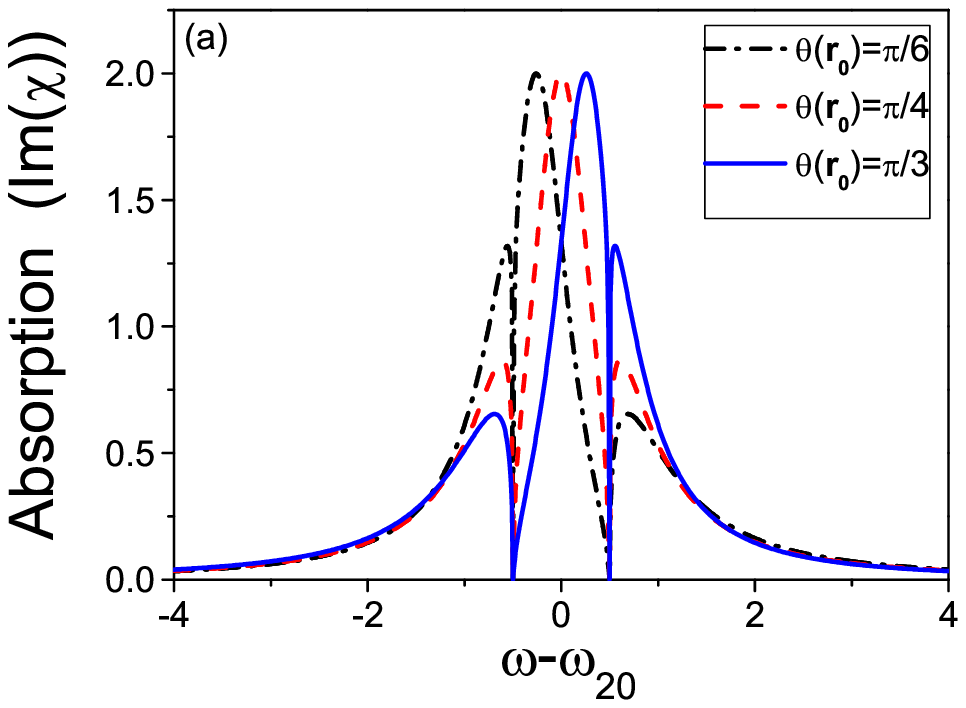} \label{Fig4a}
\includegraphics[width=12.0 cm]{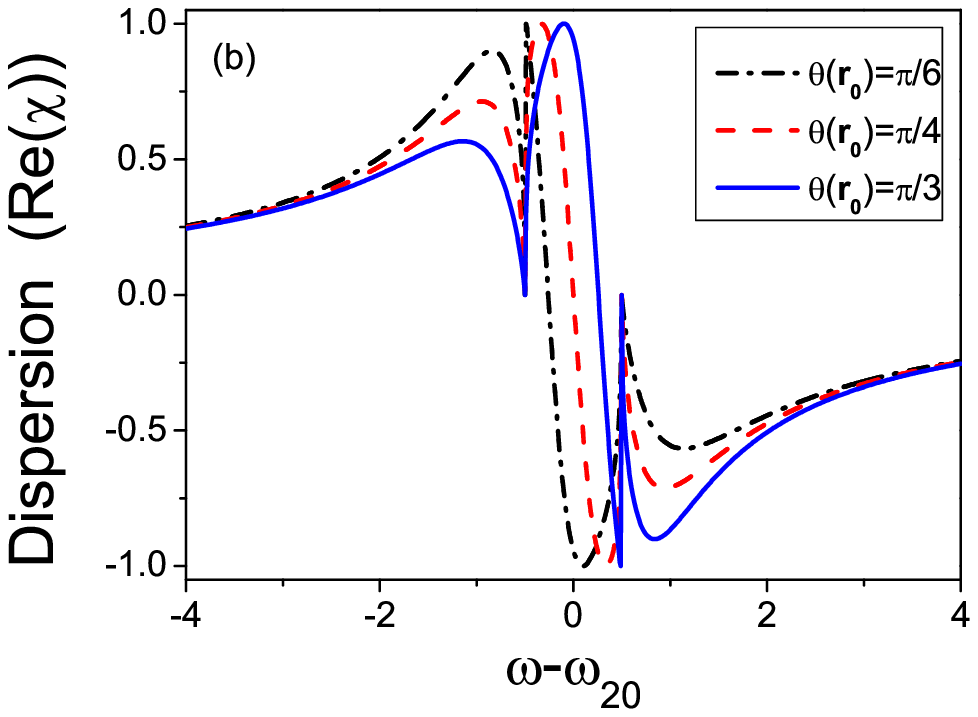} \label{Fig4b}
\caption{ (a) Absorption and (b) dispersion of the system. The parameters are $\gamma$=1, $\delta_{c}$=-$\delta_{v}$=0.5 (symmetry case), $\Delta$=1 (narrow gap), and $\theta(\vec{r}_{0})$=$\pi$/6 (dotted line); $\theta(\vec{r}_{0})$=$\pi$/4 (dashed
line); $\theta(\vec{r_{0}})$=$\pi$/3 (solid line).}%
\end{figure}
 

\end{document}